\documentclass[showkeys,showpacs,preprintnumbers,aps,twocolumn]{revtex4}

\usepackage{bm}
\usepackage[english]{babel}
\usepackage[latin1]{inputenc}
\usepackage[dvips]{graphicx}
\usepackage{amsmath}
\usepackage{epsfig}

\begin{document}
\parindent = 0pt


\date{\today}
\title{Coulomb Effects in Tunneling through a Quantum Dot Stack}

\author{H.~Sprekeler}
\author{G.~Kie{\ss}lich} 
\email{kiesslich@physik.tu-berlin.de} 
\thanks{Fax: +49-(0)30-314-21130}
\author{A.~Wacker}
\author{E.~Sch{\"o}ll} 
\affiliation{Institut f{\"u}r Theoretische Physik,
Technische Universit{\"a}t Berlin, D-10623 Berlin, Germany}

\begin{abstract}

Tunneling through two vertically coupled quantum dots is studied by means of a
Pauli master equation model. 
The observation
of double peaks in the current-voltage characteristic in a recent experiment
is analyzed in terms of the tunnel coupling between the quantum
dots and the coupling to the contacts.  Different regimes for the emitter
chemical potential indicating different peak  scenarios in the tunneling
current are discussed in detail.
We show by comparison with a density matrix
approach that the interplay of coherent and incoherent effects  in the
stationary current can be fully described by this approach.  
\end{abstract}

\maketitle

\section{Introduction}

Resonant tunneling through coupled quantum dots (QDs) is an active field of
theoretical and experimental  research (for a recent review see \cite{WIE03}).
Various nano-scale geometries for double QD systems were under investigation:
lateral gate-defined devices  (e.g. in Ref.~\cite{VAA95}) or vertical mesa
structures as proposed in Ref.~\cite{KLI94}.  With the recent advances in
fabrication of stacked  self-organized QDs where only few stacks can be
selected by electronic transport measurements an ideal system for studying
``artificial molecules''  was developed \cite{BOR01,BRY02,BRY03}. The
current-voltage  characteristic of such a double QD stack can show  peaks with
very high peak-to-valley ratio \cite{BOR01} due to the sharp energetic
resonance condition of levels in different QDs.  In Ref.~\cite{BOR01} the
authors also report on the frequent observation of double peaks in the
tunneling current which they relate to the effect of  Coulomb interaction. In
the present paper, we give a detailed theoretical examination of this
experimental feature and we analyze the mechanisms of the  current peak
broadening and height.  We use a Pauli master equation model where the
tunneling to the contacts is treated by tunneling rates and the coupling
between the QDs is described  by Fermi's golden rule.  The result of this
treatment for the stationary current through noninteracting QDs is in
agreement with recent density matrix calculations.  We will show that the
occurrence of double peaks in the current depends on the emitter chemical
potential, on the tunneling rates, and on the Coulomb interaction strength of
electrons inside the QD as well as in different QDs. Different scenarios
corresponding to different parameter regimes will be considered.

The outline of the paper is as follows: in Sec.~\ref{sec:model} we introduce
the model based on  a Pauli master equation; in Sec.~\ref{sec:disc} we present
the analytical and numerical results for the stationary current through the
double QD stack, followed by a discussion. In particular,  this section
is subdivided into the case of noninteracting electrons (Sec.~\ref{sec:noninter})
where we discuss the effect of current  peak broadening and height, and into
the case of Coulomb interacting electrons (Sec.~\ref{sec:inter})  where double
current peaks under certain conditions occur; conclusions are drawn in
Sec.~\ref{sec:conclude}.

\section{Model}
\label{sec:model}

\begin{figure}[htb]
  \begin{center}
    \includegraphics[scale=.5]{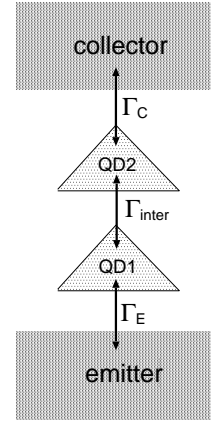}
    \caption{Scheme of the quantum dot stack connected to emitter/collector 
      contact with tunneling rates 
      $\Gamma_{E/C}$, respectively, and tunneling rate
      $\Gamma_{\mathrm{inter}}$ between the quantum dots.}
    \label{fig1}
  \end{center}
\end{figure}

In Fig.~\ref{fig1} the double QD system is sketched: QD1 is connected to an
emitter contact with a tunneling rate $\Gamma_E$, QD2 is connected to a
collector contact with a  tunneling rate  $\Gamma_C$, and both QDs are tunnel
coupled mutually with a rate $\Gamma_{\mathrm{inter}}$ (for the definition see
below).  We consider one spin degenerate single-particle state per QD.  The
many-particle (Fock-)  states $\nu$ of the QD stack are characterized by
occupation numbers $n_{is}^{(\nu)}\in\{0,1\}$, where $i=1,2$  
labels the QD and $s=\uparrow,\downarrow$ labels
the spin degree of freedom. Thus $\nu
=(n_{1\uparrow}^{(\nu)},n_{1\downarrow}^{(\nu)},
n_{2\uparrow}^{(\nu)},n_{2\downarrow}^{(\nu)})$ gives 16 different 
Fock states.

The emitter/collector contact is assumed to be in local equilibrium
characterized by Fermi functions 
$n_F^{E/C}(E)=1/(1+\exp((E-\mu_{E/C})/k_BT))$  with chemical potentials
$\mu_{E/C}$, respectively, and  temperature $T$.  An applied bias voltage
$V=(\mu_E-\mu_C)/e$ drives the QD system out of equilibrium.  Throughout the
calculations we choose the sign of the elementary charge $e>0$ so that $V>0$
means $\mu_E>\mu_C$ and $I>0$ describes a net particle current from emitter to
collector. The single-particle energies $E_i$ in the QDs are assumed to shift
linearly  with respect to  the bias voltage $V$: $E_{i}(V)=E_{i}-e\eta_iV$
with the leverage factor $\eta_i$ for the $i$-th QD.  The total energy of the
state $\nu$ is

\begin{equation}
E(\nu )=\sum_{i=1}^2N_i(\nu )(E_i-e\eta_iV)+U(\nu )
\label{eq:energy}
\end{equation}

where $N_i(\nu )=n_{i\uparrow}^{(\nu)}+n_{i\downarrow}^{(\nu)}$ 
is the particle number of the $i$-th QD for state $\nu$ and
$U(\nu )$ is the charging  energy for state $\nu$ which is given by
\cite{BEE91a,KIE02,KIE02a,KIE03b}

\begin{multline}
U(\nu )=\frac{e^2}{2}\sum_{i,j}N_i(\nu)\left[C^{-1}\right]_{ij}N_{j}(\nu)\\
-\frac{e^2}{2}\sum_{i}\left[C^{-1}\right]_{ii}N_i(\nu)
\label{eq:coulomb}
\end{multline}

with the inverse of the capacitance matrix $C$, where the last term
subtracts the self-interaction.

The evolution of the system is described by the
Pauli master equation

\begin{equation}
\frac{d}{dt} \underline{P}=\underline{\underline{M}}\cdot\underline{P}
\label{eq:evolution}
\end{equation}

for the occupation  probabilities $\underline{P}=(\{P_\nu\})^T$.
The matrix $\underline{\underline{M}}$
contains the  transition rates $M_{\nu\nu'}$ between the 
Fock states $\nu'\to \nu$.
Confining attention to single particle transitions all elements $M_{\nu\nu'}$ are zero if the
total particle number $N(\nu )=\sum_i N_{i}(\nu)$ differs by more than one
from $N(\nu' )$.
For a transition which increases the total particle number by unity, i.e. an
electron enters the QD stack via  the emitter/collector barrier, respectively,
the transition rate reads  $M_{\nu\nu'}=\Gamma_{E/C}n_F^{E/C}(\Delta
E_{\nu\nu'})$ where $\Delta E_{\nu\nu'}=E(\nu )-E(\nu')$  is the difference 
of total energies.
For the inverse tunneling processes the
transition rate is  $M_{\nu\nu'}=\Gamma_{E/C}(1-n_F^{E/C}(\Delta
E_{\nu'\nu})$). 
Transitions which do not change the total particle  number describe
tunneling between QD1 and QD2: 
Taking into account the broadening of the transition due to the
finite lifetime in the quantum dots, Fermi's golden rule 
provides us with 

\begin{equation} 
\begin{split}
M_{\nu\nu'}
=&2\pi\vert \Omega\vert^2\left[\frac{1}{\pi}\frac{\Gamma
/2}{(\Delta E_{\nu\nu'})^2 +(\Gamma /2)^2}\right]\\
=&
\Gamma_{\mathrm{inter}}\, L(\Delta E_{\nu\nu'},\Gamma)
\label{eq:golden}
\end{split}\end{equation}
Here $\Omega$ is the tunneling matrix element and
$\Gamma\equiv\Gamma_E+\Gamma_C$. The maximum tunneling rate is 
$\Gamma_{\mathrm{inter}}=4\vert\Omega\vert^2/\Gamma$
and $L(x,w)\equiv(1+(2x/w)^2)^{-1}$  is a Lorentzian function in $x$ with
a FWHM of $w$. For convenience we measure rates and energies in
the same units here, i.e. setting $\hbar =1$.

Throughout this paper we investigate the
steady state current so that the stationary occupation  probabilities
$P_\nu^0$ evaluated by $\underline{\underline{M}}\cdot\underline{P}^0=0$ 
are used. In this case the current can be evaluated
by counting the tunneling events over the collector barrier.
Summing the corresponding transitions probabilities we find
\begin{equation}
I=\sum_{\nu,\nu'}eM_{\nu'\nu}\left(1-\delta_{N(\nu )N(\nu')}\right)(N_2(\nu
)-N_2(\nu'))P_\nu^0
\label{eq:current}
\end{equation}
where
the additional factors ensure that the tunneling
process occurs at the  collector barrier and the correct sign is taken
into account for the
current direction.

\section{Discussion and Results}
\label{sec:disc}

\subsection{Noninteracting quantum dots: Current peak width and height}
\label{sec:noninter}

If the energies of single-particle states in two different QDs are resonant, a
current flows through the QD stack and a peak arises in the
current-voltage characteristic.  In this section we give an analytical
derivation of the current for noninteracting QDs, i.e.  
setting $U(\nu)\equiv 0$ in Eq.~(\ref{eq:energy}).
Then both spin directions decouple and we can restrict ourselves
to a single spin direction here. For the single-particle
levels in QD1 and QD2 it is assumed that  $E_1-\mu_E\gg k_BT$  and
$\mu_C-E_2\gg k_BT$ such that $n_F^E\approx$1 and $n_F^C\approx$0. 
With 
$\underline{P}^0=(P_{(0,0)}^0 ,P_{(1,0)}^0 ,P_{(0,1)}^0 ,P_{(1,1)}^0 )^T$
the matrix  $\underline{\underline{M}}$ simply reads

\begin{eqnarray}
\underline{\underline{M}}= \left(
\begin{array}{cccc}
-\Gamma_E & \Gamma_C & 0 & 0\\ 0 & -(\Gamma_E+\Gamma_C+Z) & Z & 0\\ \Gamma_E &
Z & -Z & \Gamma_C\\ 0 & \Gamma_E & 0 & -\Gamma_C
\end{array}
\right)
\label{eq:matrix}
\end{eqnarray}

with $Z\equiv\Gamma_{\mathrm{inter}}L(\Delta E_{(1,0)(0,1)},\Gamma)$. 
The stationary
master equation can be rewritten in a rate equation for the occupation numbers
$n_1$ and $n_2$ which results in the following stationary solution

\begin{eqnarray}
\left(
\begin{array}{c}
n_1\\ n_2
\end{array}
\right)= \frac{\Gamma_E}{\Gamma_E\Gamma_C+\Gamma Z} \left(
\begin{array}{c}
\Gamma_C+Z\\ Z
\end{array}
\right)
\label{eq:occup}
\end{eqnarray}

The current (\ref{eq:current}) then becomes

\begin{equation}
I=e\left[\frac{1}{\Gamma_E}+\frac{1}{\Gamma_C}+\frac{1}{Z}\right]^{-1}
\label{eq:current-rate}
\end{equation}

which can be physically interpreted quite easily: the three barriers  in the
QD stack act as a series connection of resistors corresponding to the
inverse of tunneling rates.  After a straightforward transformation
of (\ref{eq:current-rate}) one obtains

 \begin{equation}
I=e\Gamma_w\, L\left(\Delta E,\Gamma \sqrt{1+\frac{4\vert
\Omega\vert^2}{\Gamma_E\Gamma_C}}\right)
\label{eq:current-rate2}
\end{equation}

with
$\Gamma_w\equiv\left[\frac{1}{\Gamma_E}+\frac{1}{\Gamma_C}+\frac{1}{\Gamma_{\mathrm{inter}}}\right]^{-1}$.
Due to the assumption of a linear dependence of the single-particle levels on
the bias voltage it  follows for the energy difference  $\Delta
E=E_1-e\eta_1V-(E_2-e\eta_2V)=e(\eta_2-\eta_1)(V-V_R)$ with the bias voltage
for the resonance  $eV_R\equiv\frac{E_2-E_1}{\eta_2-\eta_1}$. Substituting
this in Eq.~(\ref{eq:current-rate2}) leads to

\begin{equation}
I(V)=e\Gamma_w\, L\left(V-V_R,\frac{\Gamma}{e(\eta_2-\eta_1)}\sqrt{1+\frac{4\vert
\Omega\vert^2} {\Gamma_E\Gamma_C}}\right)
\label{eq:current-rate3}
\end{equation}

As expected, the current-voltage characteristic shows a Lorentzian peak at the
bias voltage $V_R$.  Interestingly, its broadening is not just given by the
tunneling coupling to the contacts $\Gamma$, but by a line width modified by
the correction factor  $\sqrt{1+\frac{4\vert
\Omega\vert^2}{\Gamma_E\Gamma_C}}$. This factor provides a
significant increase of the FWHM for the current peak
if the tunneling matrix element $\vert\Omega\vert$  is larger
than the geometric mean $\sqrt{\Gamma_E\Gamma_C}$ of the
contact tunneling rates.

The expression (\ref{eq:current-rate2}) for the stationary current was also
found in  \cite{GUR96c,GUR98,WEG99} where a density matrix approach was used.
In the Appendix A we show that both approaches give
identical results in the stationary state if electron-electron
interaction is neglected.

To discuss Eq.~(\ref{eq:current-rate3}) let us assume that
$\Gamma_C>\vert\Omega\vert >\Gamma_E$: the current is limited by the lowest
rate $\Gamma_{\mathrm{inter}}L(\Delta E,\Gamma)$ for bias voltages $V$ far
away from $V_R$. If the bias approaches $V_R$ 
this rate is increased
until it equals the emitter coupling $\Gamma_E$. Now, the emitter barrier
starts to limit the current, i.e.  the current flattens close to the resonance
so that the FWHM value of the current is reached for larger distances from the
resonance voltage $V_R$.  The smaller the ratio of 
$\Gamma_E/\vert\Omega\vert$ the broader the current peak.  

The peak (maximum) current
is given by $I(V_R)=e\Gamma_w$.  It has a non-monotonic dependence on the
collector tunneling rate $\Gamma_C$:  for weak collector coupling the peak
current increases linearly, it shows a maximum at $ \Gamma_C=2|\Omega|$
and drops as $1/\Gamma_C$ for  strong collector coupling. (The same
holds for the emitter coupling.)
On the basis of the coherent description
this was interpreted in Ref.~\cite{WEG99}
as a competition between enhanced tunneling for weak collector  
coupling and destructive
interference in the opposite limit.  Hence, it is worth to note that one
obtains the same effect in a Fermi's golden rule treatment 
where no interference is taken into account for. In this picture
increasing 
$\Gamma_C$ enhances the transport over the collector barrier but
limits the transport between the dots by broadening the transition.

\subsection{Coulomb interacting quantum dots}
\label{sec:inter}

Now, our considerations will be extended to the realistic situation of Coulomb
interacting electrons. We describe the interaction of 
electrons inside QD1 and
QD2 with charging energies $e^2[C^{-1}]_{11}=U_1$ and $e^2[C^{-1}]_{22}=U_2$, 
respectively.  The Coulomb
interaction strength of electrons in different QDs is given by
$e^2[C^{-1}]_{12}=U_{\mathrm{inter}}$.

\subsubsection{Dependence on emitter Fermi energy}

\begin{figure}[htb]
  \begin{center}
    \includegraphics[width=.4\textwidth]{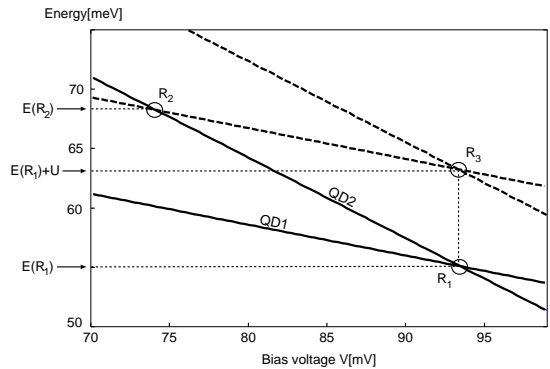}
    \caption{Linear bias voltage dependence of the single-particle 
      energy levels in QD1 ($\eta_1=$0.26, $E_1=$79.5 meV) and in 
      QD2 ($\eta_2=$0.68, $E_2=$118.7 meV).  
      Dashed lines: addition energies for the
      second electron ($U=U_1=U_2=$8meV,  $U_{\mathrm{inter}}=$0).}
    \label{fig2}
  \end{center}
\end{figure}

In Fig.~\ref{fig2} the bias voltage dependence of the QD energies are
depicted.  The zero-bias single-particle energies $E_i$
and leverage factors $\eta_i$  were
obtained from the calculation of the transmission
through a two-dimensional geometry 
as depicted in Fig.~\ref{fig1} without scattering processes \cite{SPR03a}.
In order to describe the experiment, 
we use the geometry in Ref.~\cite{BRY03} which gives 
$\eta_1=$0.26, $E_1=$79.5 meV, and $\eta_2=$0.68, $E_2=$118.7 meV.

Due to the different slopes, the  single-particle levels of QD1 and QD2 (full
lines in Fig.~\ref{fig2}) intersect at the point $R_1$.  Further resonances
can occur by considering the double occupancy of the QDs. Assuming
equal charging energies for both  QDs ($U=U_1=U_2=$ 8meV) parallel lines with
the additional charging energy $U$ emerge (dashed lines in Fig.~\ref{fig2}).
The resonance $R_2$ is due to the intersection of the energies of the doubly
occupied QD1 and the singly occupied QD2. The  resonance of the doubly
occupied states $R_3$  occurs at the same bias voltage as for $R_1$ 
if the
charging energies for both QDs are equal.

\begin{figure}[htb]
  \begin{center}
    \includegraphics[width=.4\textwidth]{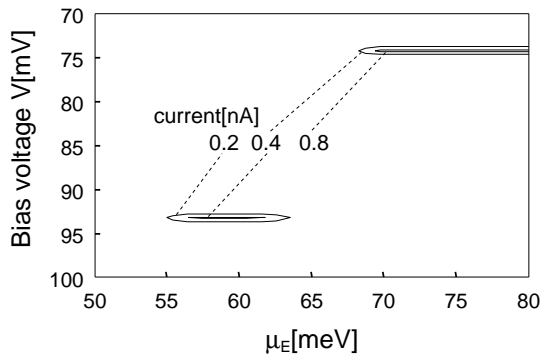}
    \caption{Contour plot: current I vs. emitter chemical potential $\mu_E$ and bias voltage 
        $V$ for $U=U_1=U_2=$ 8meV, $U_{\mathrm{inter}}=$ 0,
        $\Gamma_E=\Gamma_C=$ 100$\mu$eV, $\vert\Omega\vert =$10$\mu$eV,
        $T=4.2$ K. (Isolines correspond to 0.2, 0.4, 0.8 nA)}
    \label{fig3}
  \end{center}
\end{figure}

A current flow through the QD system is possible at the resonance energies
$E(R_1)$ and $E(R_2)$  providing that electron states in the emitter contact
are occupied at these energies. (Typically, 
these resonances occur at energies far above the chemical potential 
of the  collector Fermi, so that the electrons can always leave the
QD system via the collector barrier.)
In Fig.~\ref{fig3} the dependence  of the
current on the emitter chemical potential  $\mu_E$ and the bias voltage $V$ is
shown as a contour plot. Here we assume
that the tunnel coupling between the dots is rather weak,
i.e. $\vert\Omega\vert\ll\Gamma_E,\Gamma_C$, so that the
occupation of the dots is determined by the respective reservoirs. 
The conduction band edge in the  emitter is the
zero-point of the energy scale.  Different scenarios for the stationary
current depending on $\mu_E$ arise:
\begin{itemize}
\item[(i)] $\mu_E<E(R_1)$: there is only an exponential small current due to
thermal activated  electrons in the emitter.
\item[(ii)] $E(R_1)<\mu_E<E(R_1)+U=E(R_3)$: in this case the current shows a
peak due to the  resonance $R_1$ of the single-particle levels, the  second
resonance $R_2$ is energetically available for thermally activated electrons
from the  emitter and gives rise to an additional current peak at lower bias
voltage which increases with increasing temperature  (this scenario was
experimentally investigated in \cite{BRY03} and was used to determine the
charging energy $U$).
\item[(iii)] $E(R_1)+U<\mu_E<E(R_2)$: no current peak appears. In this regime
it is  possible to add a second electron in QD1. Since the coupling between
the QDs is weaker than the coupling to the emitter contact, QD1 is mostly
occupied with two electrons. They leave QD1 with the energy $E_1+U$ so that
they cannot fulfill the resonance condition for $R_1$.
\item[(iv)] $\mu_E>E(R_2)$: a current peak due to the resonance $R_2$ occurs
with twice the  peak height than in the case (ii) because two electrons
contribute to the current here. A current peak for resonance $R_1$ is
suppressed for the same reasons as in (iii).
\end{itemize}

\subsubsection{Dependence on tunnel coupling}

Now we want to study the behavior with increasing tunnel coupling 
$\vert\Omega\vert$. In particular
we consider the  regime (iv) of the last section but now  we allow the
tunneling rates to the contacts to be different  ($\Gamma_E=$ 17$\mu$eV,
$\Gamma_C=$ 400$\mu$eV).

\begin{figure}[htb]
  \begin{center}
    \includegraphics[width=.4\textwidth]{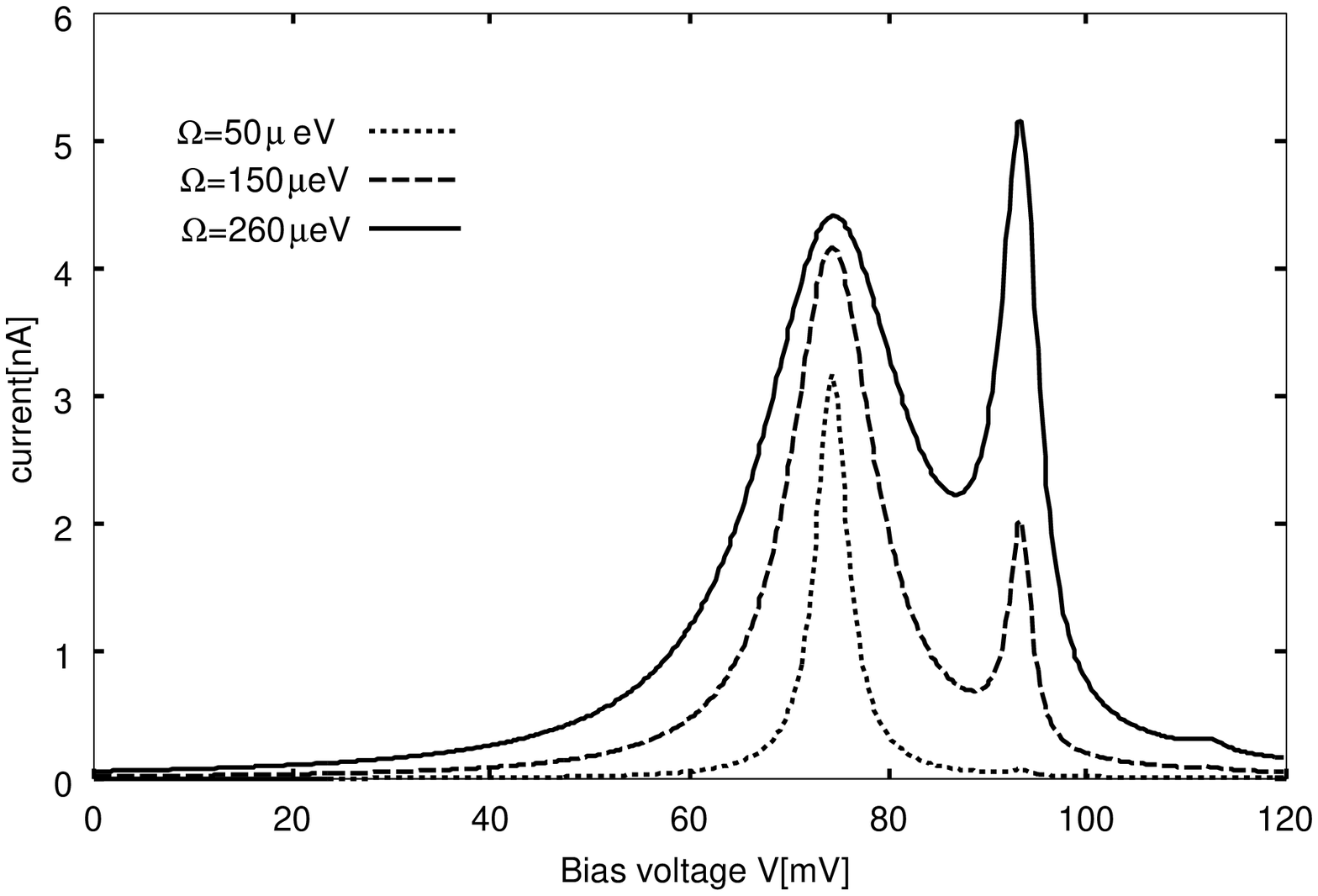}
    \caption{Current I vs. bias voltage $V$ for varying coupling between QDs 
        $\vert\Omega\vert$. $\Gamma_E=$ 17$\mu$eV, $\Gamma_C=$ 400$\mu$eV,
    $T=$ 4.2K, $U=U_1=U_2=$ 8meV, $U_{\mathrm{inter}}=$ 0, $\mu_E=$ 90meV}
    \label{fig4}
  \end{center}
\end{figure}

Fig.~\ref{fig4} shows the bias voltage dependent current for varying couplings
$\vert\Omega\vert$  between the QDs. For low $\vert\Omega\vert$ one observes
a current
peak at $V=$ 75mV corresponding to the resonance $R_2$ in
Fig.~\ref{fig2} (compare regime (iv) in the last section). According to
Eq.~(\ref{eq:current-rate3})  the current peak broadens with increasing
$\vert\Omega\vert$ but one has to consider the double  occupancy of QD1 which
modifies the current peak width such that the correction factor becomes now
$\sqrt{1+2\cdot\frac{4\vert\Omega\vert^2}{\Gamma_E\Gamma_C}}$.  With
increasing $\vert\Omega\vert$ a second current peak at the bias voltage of
$R_1$ appears in Fig.~\ref{fig4}.  Due to the stronger coupling between the
QDs single occupancy of QD1  becomes more likely, since the second electron in
QD1 can tunnel into  QD2 off-resonantly because of the finite linewidth of the
levels.  Hence, electrons  with energy $E_1$ can leave QD1 and contribute in
resonance $R_1$ which generates a current peak at $V(R_1)$. Hence, the
possibility of the  observation of a double current peak in the experiment
depends mainly on the tunnel matrix element $\Omega$ between states in
different QDs.  Its sensitivity on the spatial separation of the QDs  within
the stack could explain the frequent 
observation of the double current peaks \cite{BOR01}.

Note that for $\vert\Omega\vert\ge$ 260$\mu$eV a small  third current peak at
almost 110 mV appears due to the resonance  of the energies of the doubly
occupied QD2 and single occupied QD1  (out of the bias range depicted in
Fig.~\ref{fig2}).

\begin{figure}[htb]
  \begin{center}
    \includegraphics[width=.4\textwidth]{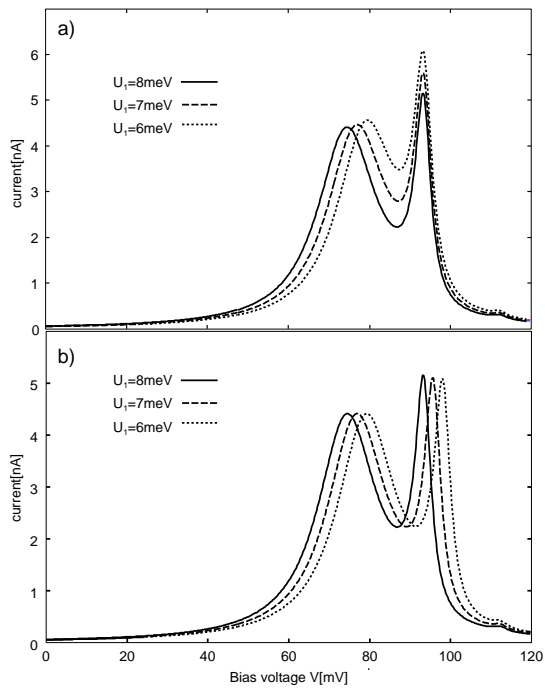}
    \caption{Current I vs. bias voltage $V$ for different charging 
      energies $U_1$ of QD1; $U_2=$ 8meV; a) $\Gamma_E=$ 17$\mu$eV, $\Gamma_C=$ 400$\mu$eV; b)
      $\Gamma_E=$ 400$\mu$eV, $\Gamma_C=$ 17$\mu$eV; other
      parameters  are the same as  for the full curve in Fig.~\ref{fig4}}
    \label{fig5}
  \end{center}
\end{figure}

Underlining the corresponding mechanisms which lead to the current peaks in
Fig.~\ref{fig4} we show the current-voltage characteristics for different
charging energies $U_1$ with fixed $U_2=$ 8meV in Fig.~\ref{fig5}a
(the other parameters are the same as for the full curve in Fig.~\ref{fig4}).
As one can see, with decreasing $U_1$ the low bias current peak shifts to higher bias
voltages and the second current peak remains at the same bias voltage position: Reducing
$U_1$ leads to a parallel shift  of the dashed line with smaller slope in
Fig.~\ref{fig2}  so that the intersection point $R_2$ shift to
higher bias voltages and the resonance $R_1$ leaves unchanged. 
In Fig.~\ref{fig5}b the values of $\Gamma_E$ and $\Gamma_C$ were interchanged so that a different
physical situation emerges. Here, both current peaks shift to higher bias voltages by the same amount 
with decreasing $U_1$, i.e. the resonance $R_3$ is now mainly responsible for the current peak at higher bias voltage. 
Due to the weaker collector coupling the double occupancy of both QDs becomes more likely.  

Note, that for equal charging energies in QD1 and QD2  the current-voltage characteristics are
invariant with respect  to an interchange of $\Gamma_E$ and $\Gamma_C$. 
This is due to the electron-hole symmetry in the considered QD system. The
necessary condition for this symmetry is that the resonances occur at energies
deep in the Fermi sea on one contact side and unoccupied states on the other
side.

\subsubsection{Coulomb interaction between quantum dots}

\begin{figure}[htb]
  \begin{center}
    \includegraphics[width=.4\textwidth]{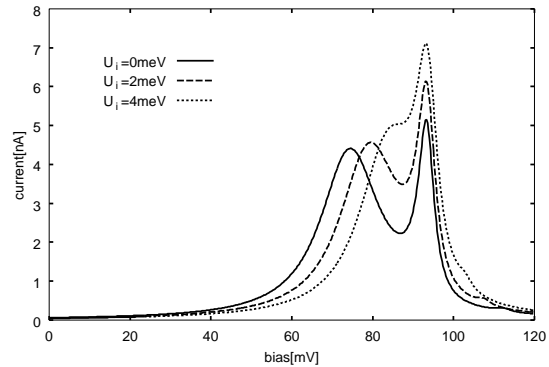}
    \caption{Current I vs. bias voltage $V$ for different Coulomb interaction strengths between 
        QD1 and QD2: $U_{\mathrm{inter}}$, other parameters  are the same as
      for the full curve in Fig.~\ref{fig4}}
    \label{fig6}
  \end{center}
\end{figure}

The self-organized QDs within one stack grown in \cite{BOR01,BRY02,BRY03} have
a spatial separation  of few nm so that the Coulomb interaction between
electrons in different QDs is a crucial ingredient in a theoretical
examination. For different interaction  strengths $U_{\mathrm{inter}}=$ 0, 2,
4meV  the current vs. bias voltage is shown in Fig.~\ref{fig6}.

Again a double peak appears whereas the distance of the maxima diminishes with
increasing  $U_{\mathrm{inter}}$.  If QD1 is filled with one electron the
single-particle level of QD2 is shifted by $U_{\mathrm{inter}}$  so that the
low-bias resonance $R_2$ moves to higher bias voltages. The resonance $R_1$
stays at the same bias voltage since QD2  filled with one electron generates a
shift  of the single-particle level of QD1 by the same amount
$U_{\mathrm{inter}}$. This gives for  the current peak separation

\begin{equation}
\Delta V=\frac{U-U_{\mathrm{inter}}}{e(\eta_2-\eta_1)}
\end{equation}

Another feature  can be seen in the dotted curve in Fig.~\ref{fig6}: if
$\Delta V$ becomes smaller, the  height of the high-bias peak increases. This
peak structure is quite similar to that observed in \cite{BOR01}.

\section{Conclusions}
\label{sec:conclude}

The non-equilibrium current-voltage characteristic of a system of two QDs
coupled in series  was investigated by a Pauli master equation approach. 
The different leverage factors of the  energy levels in the QDs
leads to resonances at certain  bias voltages.
If the emitter contact provides
electrons at the resonances 
a current through the QD system is driven. The height and  the width
of the arising current peaks in the current voltage characteristic was
examined analytically for the case of
vanishing Coulomb interaction. Here it was found that the
width of the resonance increases if the tunnel coupling between the QDs
surpasses the geometric mean of the couplings to the contacts. 
These results are in full agreement with a fully coherent approach.
Including the Coulomb interaction
a double current peak can arise for
asymmetric coupling to the contacts and a tunnel coupling of the order of the
Coulomb interaction strengths. The strong sensitivity of  the coupling
parameters to the barrier thicknesses in self-organized QD stacks can
provide a possible explanation of the frequent observation of  double current
peaks in the experiment.

\begin{acknowledgments}
The authors would like to thank T. Bryllert for helpful discussions.  This
work was supported by Deutsche Forschungsgemeinschaft in the framework of Sfb
296.
\end{acknowledgments}

\appendix
\section{Derivation of the master equation from density matrix approach}

We start with the modified Liouville equation for the double QD system
(Fig.~\ref{fig1}) derived in Ref.~\cite{GUR96c}.  The following abbreviations
for the Fock states $\vert\nu\rangle$ will be used: $\vert a\rangle\equiv
\vert 0,0\rangle$,  $\vert b\rangle\equiv \vert 1,0\rangle$,  $\vert
c\rangle\equiv \vert 0,1\rangle$, and $\vert d\rangle\equiv \vert
1,1\rangle$. Then the time evolution of  the corresponding density-matrix
elements is given by

\begin{eqnarray}
\dot{\rho}_{aa}&=&-\Gamma_E\rho_{aa}+\Gamma_C\rho_{cc}\nonumber\\
\dot{\rho}_{bb}&=&\Gamma_E\rho_{aa}+\Gamma_C\rho_{dd}+i\Omega
(\rho_{bc}-\rho_{cb})\nonumber\\ \dot{\rho}_{cc}&=&-\Gamma\rho_{cc}-i\Omega
(\rho_{bc}-\rho_{cb})\nonumber\\
\dot{\rho}_{dd}&=&-\Gamma_C\rho_{dd}+\Gamma_E\rho_{cc}\nonumber\\
\dot{\rho}_{bc}&=&i\Delta E\rho_{bc}+i\Omega
(\rho_{bb}-\rho_{cc})-\frac{1}{2}\Gamma\rho_{bc}
\label{eq:liouville}
\end{eqnarray}

with $\Delta E\equiv E_2-E_1$ and $\Gamma\equiv\Gamma_E+\Gamma_C$.
Additionally, probability conservation is demanded: $\sum_\nu\rho_{\nu\nu}=1$
and  $\rho_{cb}=\rho_{bc}^*$ holds.  Eq.~(\ref{eq:liouville}) can be transformed
in a matrix equation of the form  $\dot{\underline{u}}=\underline{\underline
A}\cdot\underline{u}$ with $\underline{u}\equiv
\left(\rho_{aa},\rho_{bb},\rho_{cc},\rho_{dd},\textrm{Re}(\rho_{bc}),
\textrm{Im}(\rho_{bc})\right)^T$ and

\begin{eqnarray}
\underline{\underline A}\equiv \left(
\begin{array}{cccccc}
-\Gamma_E & \Gamma_C & 0 & 0 & 0 & 0\\ 0 & -\Gamma & 0 & 0 & 0 & -2\Omega \\
\Gamma_E & 0 & 0 & \Gamma_C & 0 & 2\Omega\\ 0 & \Gamma_E & 0 & -\Gamma_C & 0 &
0\\ 0 & 0 & 0 & 0 & -\frac{1}{2}\Gamma & -\Delta E\\ 0 & \Omega & -\Omega & 0
& \Delta E & -\frac{1}{2}\Gamma
\end{array}
\right)
\label{eq:koeff-matrix}
\end{eqnarray}

In order to derive the Pauli master equation for the occupation probabilities
$P_\nu =\rho_{\nu\nu}$ one has to get rid of the non-diagonal elements 
$\rho_{bc}$ in
(\ref{eq:liouville}) by truncating the  6$\times$6 matrix
(\ref{eq:koeff-matrix}) to the 4$\times$4 matrix (\ref{eq:matrix}).  
This can be accomplished 
by setting the time derivatives of $\textrm{Re}(\rho_{bc})$ and
$\textrm{Im}(\rho_{bc})$ to zero.
Then one can solve the algebraic equations for
$\rho_{bc}^{\mathrm{stat}}=\Omega(\rho_{cc}-\rho_{bb})/(\Delta E+i\Gamma/2)$
and substitute this expression in
the equations for $\dot{\rho}_{bb}$ and $\dot{\rho}_{cc}$. This immediately
leads to the terms  $Z=\frac{\Omega^2\Gamma}{\Delta E^2+(\Gamma /2)^2}$  in
(\ref{eq:matrix}) as introduced by Fermi's golden rule in
Sec.~\ref{sec:model}. 

The necessary assumption $\dot{\rho}_{bc}=0$ is justified if one
of two conditions is met:
(i) In the limiting case  $\Omega\ll\Gamma$ the relaxation 
of $\rho_{bc}(t)$ to $\rho_{bc}^{\mathrm{stat}}$ occurs on a fast 
time scale. This is the condition for sequential tunneling
which underlies our derivation of the Pauli master equation.
(ii) In the stationary case $\dot{\rho}_{bc}=0$ holds
independently of the magnitude of $\Omega$. Therefore,
the Pauli master equation provides reliable results also 
in the strong coupling limit if one restricts oneself to the stationary
current.

\end{document}